\documentclass[preprint,12pt]{elsarticle}

\usepackage{amssymb}

\usepackage{graphicx}
\usepackage{wrapfig}
\usepackage{url}
\usepackage{here}
\usepackage{tabularx}
\usepackage{amsmath}
\usepackage{threeparttable}
\usepackage{siunitx}
\usepackage{multirow}

\journal{Nuclear Instruments and Methods in Physics Research, Section A (NIMA)}

\begin{document}

\begin{frontmatter}

\title{X-ray imaging camera using INTPIX4NA SOIPIX detector with SiTCP-XG 10GbE based high-speed readout system}

\author[KEKIMSS,SOKENDAI]{Ryutaro NISHIMURA\corref{mycorrespondingauthor}}
\ead{ryutaro.nishimura@kek.jp}

\author[KEKIMSS,SOKENDAI]{Noriyuki IGARASHI}

\author[KEKIMSS,SOKENDAI]{Daisuke WAKABAYASHI}

\author[KEKIMSS,SOKENDAI]{Yuki SHIBAZAKI}

\author[KEKIMSS]{Yoshio SUZUKI}

\author[KEKIMSS,SOKENDAI,TSUKUBAUNIV]{Keiichi HIRANO}

\author[KEKIPNS]{Yasuo ARAI}

\cortext[mycorrespondingauthor]{Corresponding author}

\address[KEKIMSS]{Photon Factory, Institute of Materials Structure Science,\\ High Energy Accelerator Research Organization (KEK-IMSS-PF),\\ Oho 1-1, Tsukuba, Ibaraki, 305-0801, Japan}
\address[SOKENDAI]{Materials Structure Science Program, Graduate Institute for Advanced Studies, Graduate University for Advanced Studies (SOKENDAI),\\ Oho 1-1, Tsukuba, Ibaraki, 305-0801, Japan}
\address[TSUKUBAUNIV]{Graduate School of Pure and Applied Sciences, University of Tsukuba,\\ Tennodai 1-1-1, Tsukuba, Ibaraki, 305-8571, Japan}
\address[KEKIPNS]{Institute of Particle and Nuclear Studies, High Energy Accelerator Research Organization (KEK-IPNS),\\ Oho 1-1, Tsukuba, Ibaraki, 305-0801, Japan}

\begin{abstract}

The silicon-on-insulator pixel (SOIPIX) detector is a unique monolithic-structure imaging device currently being developed by the SOIPIX group led by the High Energy Accelerator Research Organization (KEK). 
The detector team at KEK Photon Factory (PF) is also developing an X-ray camera using INTPIX4NA with a 14.1 \si{\times} 8.7 \si{mm^2} sensitive area and 425,984 (832 column \si{\times} 512 row matrix) pixels, with a pixel size of 17 \si{\times} 17 \si{\micro\meter^2}. 
The detector has high resolution and sensitivity for low-intensity X-rays, making it suitable for imaging in optical systems with lower X-ray intensities, such as an X-ray zooming microscope using two Fresnel zone plates (FZPs), which is also under development at PF. 
To enable imaging under such conditions, we developed a detector cooling system using a Peltier element to support longer exposure time (~0.5 seconds per frame). 
Additionally, we developed a new readout system using DAQ boards developed by PF, equipped with SiTCP-XG (network controller implemented on field-programmable gate array) that supports 10 Gbps Ethernet for high-frame-rate imaging at several hundred hertz. 
The new X-ray camera was tested at the PF BL-14A, BL-14B, and AR-NE1A experimental stations, and the resolution and sensitivity characteristics were confirmed. 
Given these confirmed characteristics, this X-ray camera is suitable for X-ray imaging using 5--20 keV X-rays under low-intensity, low-contrast conditions. 
These conditions are ideal for capturing soft tissues with poor contrast, objects with fine structures, and specimens vulnerable to radiation damage. 

\end{abstract}


\end{frontmatter}


\section{Introduction}
X-ray imaging is one of the most effective techniques for evaluating material structures without causing destruction. 
In particular, X-ray Schlieren microscopy is a useful method for assessing the microstructure of specimens made up of light elements, such as organism samples, utilizing X-ray absorption and phase contrast images. 
The X-ray optics team at the High Energy Accelerator Research Organization - Photon Factory (KEK-PF) is developing variable-magnitude Schlieren microscopy optics using two Fresnel zone plates (FZPs) \cite{schmic1}. 
These optics are more compact and flexible than the conventional optics \cite{schmic2}. 
However, the X-ray intensity is significantly reduced by the two FZPs, emphasizing the necessity for a detector with high sensitivity and resolution characteristics. 

We chose INTPIX4NA \cite{i4na}, a silicon-on-insulator pixel (SOIPIX) detector \cite{soi1, soi2}, as a suitable detector for use under low X-ray intensity conditions. 
Compared with the INTPIX4NA and other detectors, the INTPIX4NA has a smaller pixel size in direct conversion-type detectors. 
The effective spatial resolution is expected to be higher than that of the indirect conversion-type detector and is anticipated to reach the upper limit constrained by pixel size (A comparison of specifications is shown in table \ref{tab:compare_detectors}). 
Based on these characteristics, we determined that the INTPIX4NA is suitable for imaging in low X-ray intensity conditions. 

\begin{threeparttable}[H]
 \centering
 \caption{Comparison of X-ray detectors}
 \begin{tabularx}{\linewidth}{XXXXX}
  & Pixel size & Pixel matrix (1 chip) & Sensitive area (1 chip) & Photon detection (conversion type) \\
 \hline
 INTPIX4NA \cite{i4na} & 17 \si{\times} 17 \si{\micro\meter^2} & 832 \si{\times} 512 & 14.1 \si{\times} 8.7 \si{mm^2} & Direct (Si) \\
 X-ray sCMOS (C12849-111U \cite{xscmos1}) & 6.5 \si{\times} 6.5 \si{\micro\meter^2} & 2048 \si{\times} 2048 & 13.3 \si{\times} 13.3 \si{mm^2} & Indirect (Gadolinium oxysulfide (P43) phosphor \SI{10}{\micro\meter}) \\
 X-ray CCD (S10810-11 \cite{xsccd1}) & 20 \si{\times} 20 \si{\micro\meter^2} & 1500 \si{\times} 1000 & 30 \si{\times} 20 \si{mm^2} & Indirect (CsI deposited on a fiber optic plate (FOP)) \\
 Eiger 2 R 500K \cite{eiger2} & 75 \si{\times} 75 \si{\micro\meter^2} & 1028 \si{\times} 512 & 77.1 \si{\times} 38.4 \si{mm^2} & Direct (Si / CdTe) \\
 Medipix 3 \cite{medipix3-1, medipix3-2} & 55 \si{\times} 55 \si{\micro\meter^2} & 256 \si{\times} 256 & 14.1 \si{\times} 14.1 \si{mm^2} & Direct (Si / CdTe) \\
 \hline
 \end{tabularx}
 \label{tab:compare_detectors}
\end{threeparttable}

To utilize this detector as an X-ray camera, a new readout system and a Peltier-based detector cooling chamber were developed. 
The new readout system is based on the SiTCP-XG field-programmable gate array (FPGA)-based 10 Gbps Ethernet network controller \cite{sitcp-xg, sitcp-xg_protodaq}. 
This new INTPIX4NA-based X-ray camera should satisfy the high sensitivity and resolution characteristics required by the two FZP optics.

\section{INTPIX4NA SOIPIX Detector}

\subsection{SOIPIX detector}

\begin{figure}[htb]
\centering
\includegraphics[width=\linewidth]{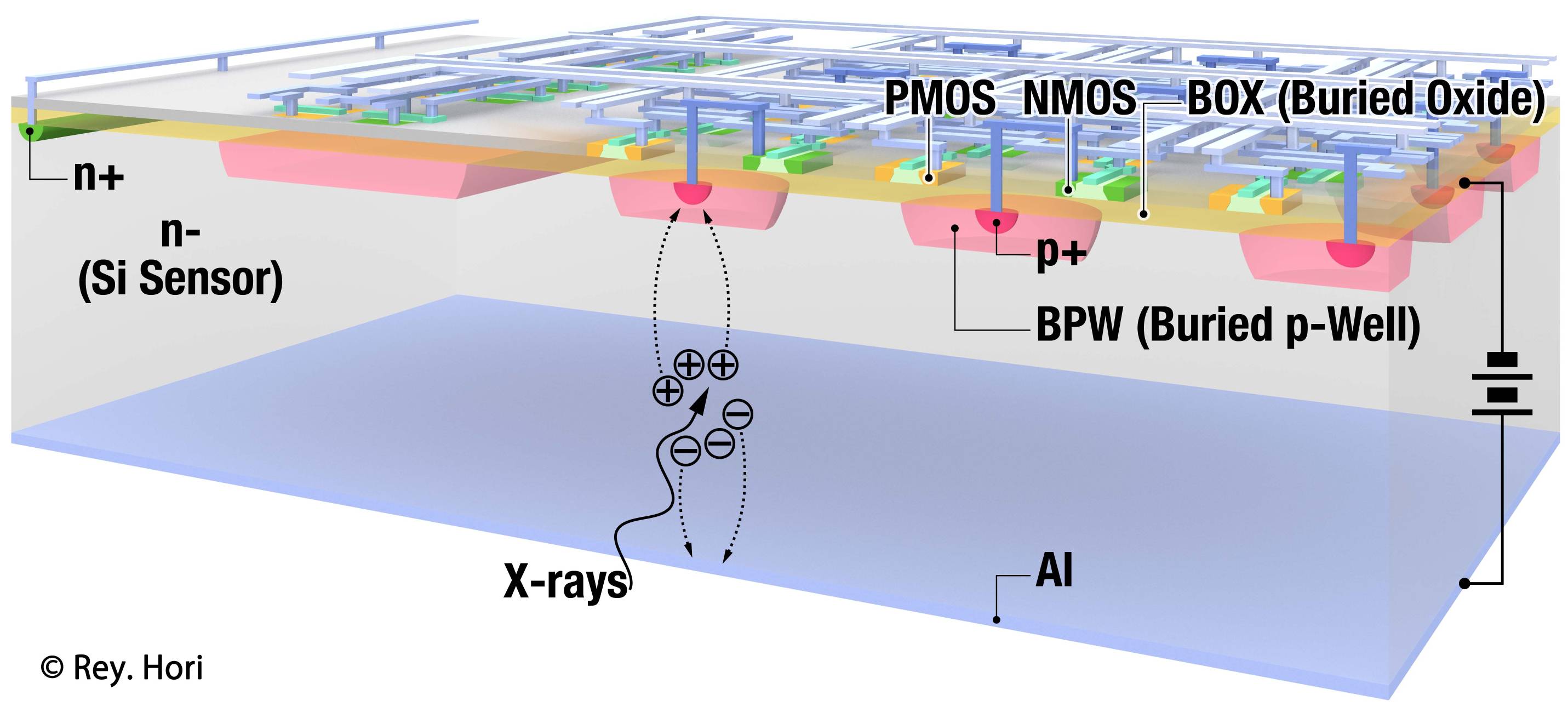}
\caption{Structure of SOIPIX detector.}
\label{fig:soistruct}
\end{figure}

SOIPIX detectors are currently being developed by the SOIPIX collaboration, led by KEK (Tsukuba, Ibaraki, Japan). 
These detectors are based on a 0.2 $\mathrm{\mu m}$ CMOS fully depleted silicon-on-insulator (FD-SOI) pixel process developed by Lapis Semiconductor Co., Ltd. \cite{soi1}. 
The SOIPIX detector comprises a thick, high-resistivity Si substrate for the sensor and a thin Si layer for the CMOS circuits (Fig. \ref{fig:soistruct}). 
Considering an SOIPIX detector has no bump bonding, the application has low parasitic capacitance (\verb|~|10 fF), low noise, high conversion gain, smaller pixel size, and low material budget. 
Furthermore, it can operate rapidly at low power. 

\subsection{Overview of INTPIX4NA}

\begin{figure}[htb]
\centering
\includegraphics[width=\linewidth]{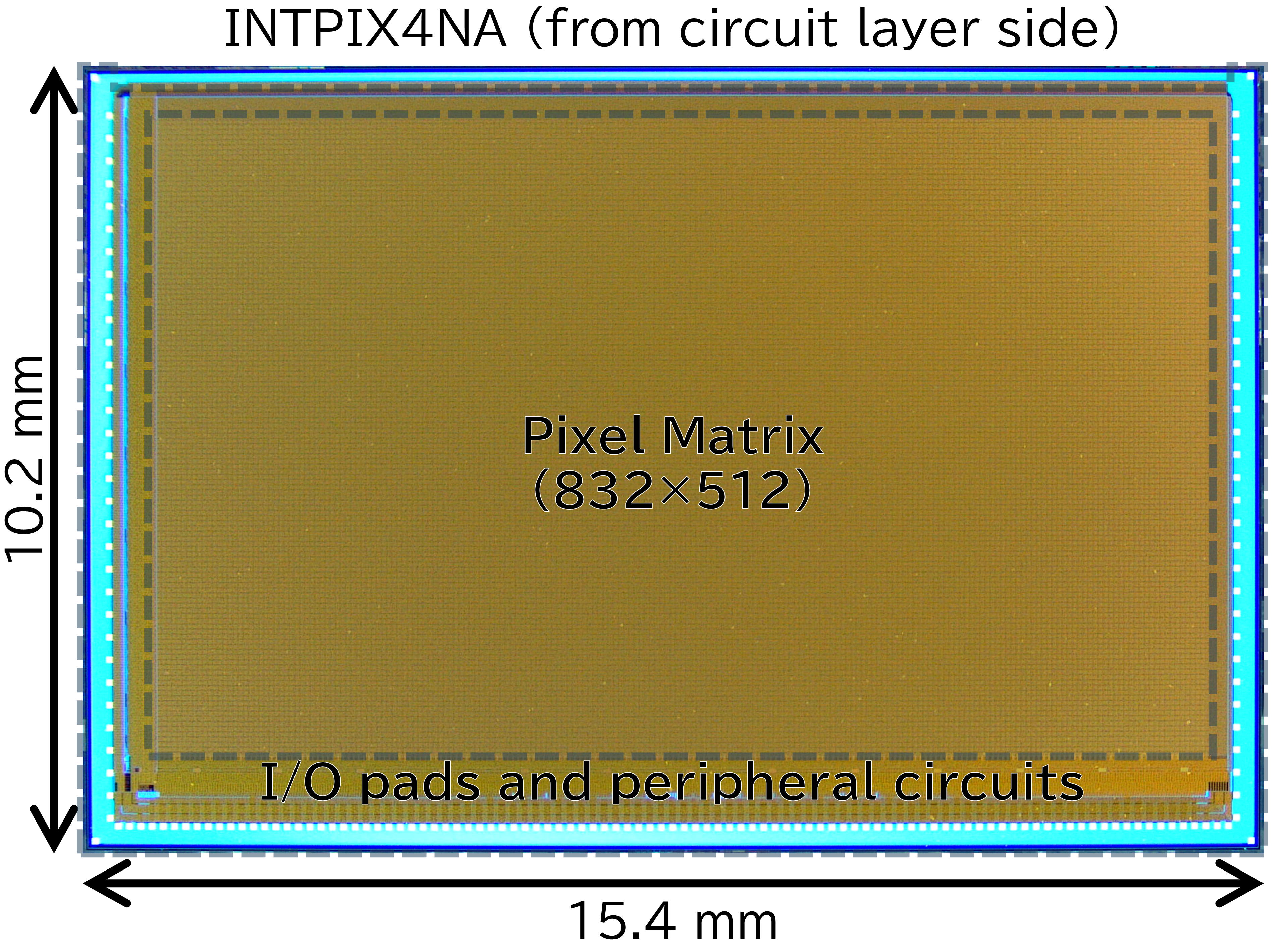}
\caption{Photograph of INTPIX4NA from circuit layer side.}
\label{fig:intpix4na}
\end{figure}

INTPIX4NA \cite{i4na} is a charge-integration-type SOIPIX detector. 
The pixel size is 17 $\times$ 17 $\mathrm{\mu m^2}$, and the number of pixels is 425,984 (832 columns $\times$ 512 rows). 
The sensitive area is 14.1 $\times$ 8.7 $\mathrm{mm^2}$. 
The detector comprises 13 blocks (64 columns $\times$ 512 rows pixels per block, and each block has independent analog output channels for a parallel readout. 
The design parameters are listed in Table \ref{tab:detector_design} and a photograph of INTPIX4NA is presented in Fig. \ref{fig:intpix4na}. 
Considering its sensitivity and spatial resolution performance from a prior study \cite{i4na}, INTPIX4NA is suitable for X-ray imaging with 5--20 keV low-intensity X-rays. 

\begin{table}[H]
 \caption{INTPIX4NA design parameters \cite{i4na}}
 \begin{tabularx}{\linewidth}{XX}
 \hline
 Chip size & 15.4 \si{\times} 10.2 \si{mm^2} \\ 
 Sensitive area & 14.1 \si{\times} 8.7 \si{mm^2} \\ 
 Pixel matrix & 832 columns \si{\times} 512 rows (\SI{425984}{pixels}) \\ 
 Pixel size & 17 \si{\times} 17 \si{\micro\meter^2} \\ 
 Pixel gain & 9.3--10.6 \si{\micro\volt/\elementarycharge} (Actual measurement value) \\ 
 Shutter & Global Shutter \\ 
 Thickness of transistor layer & \SI{40} {nm} \\ 
 Thickness of buried oxide (BOX) layer & \SI{140} {nm} \\ 
 Thickness of sensor layer & \SI{300} {\micro\meter} (Typical) \\ 
 Sensor layer wafer & N type Floating Zone wafer \\ 
 Back side Aluminum deposition thickness & \SI{150} {nm} (Typical) \\ 
 Readout mode & Full array one channel serial readout / 13 channel parallel readout (32,768 (64 columns \si{\times} 512 rows matrix) pixels/channel) \\ 
 Number of signal lines & 47 (27 digital inputs, 4 analog inputs, and 16 analog outputs) \\ 
 Power Consumption & \SI{280} {mW} (Reset state) \\ 
 Other features & in-pixel correlated double sampling (CDS) circuit is implemented. \\ 
 \hline
 \end{tabularx}
 \label{tab:detector_design}
\end{table}

\subsection{INTPIX4NA's implementation for X-ray imaging}

\begin{figure}[htb]
\centering
\includegraphics[width=\linewidth]{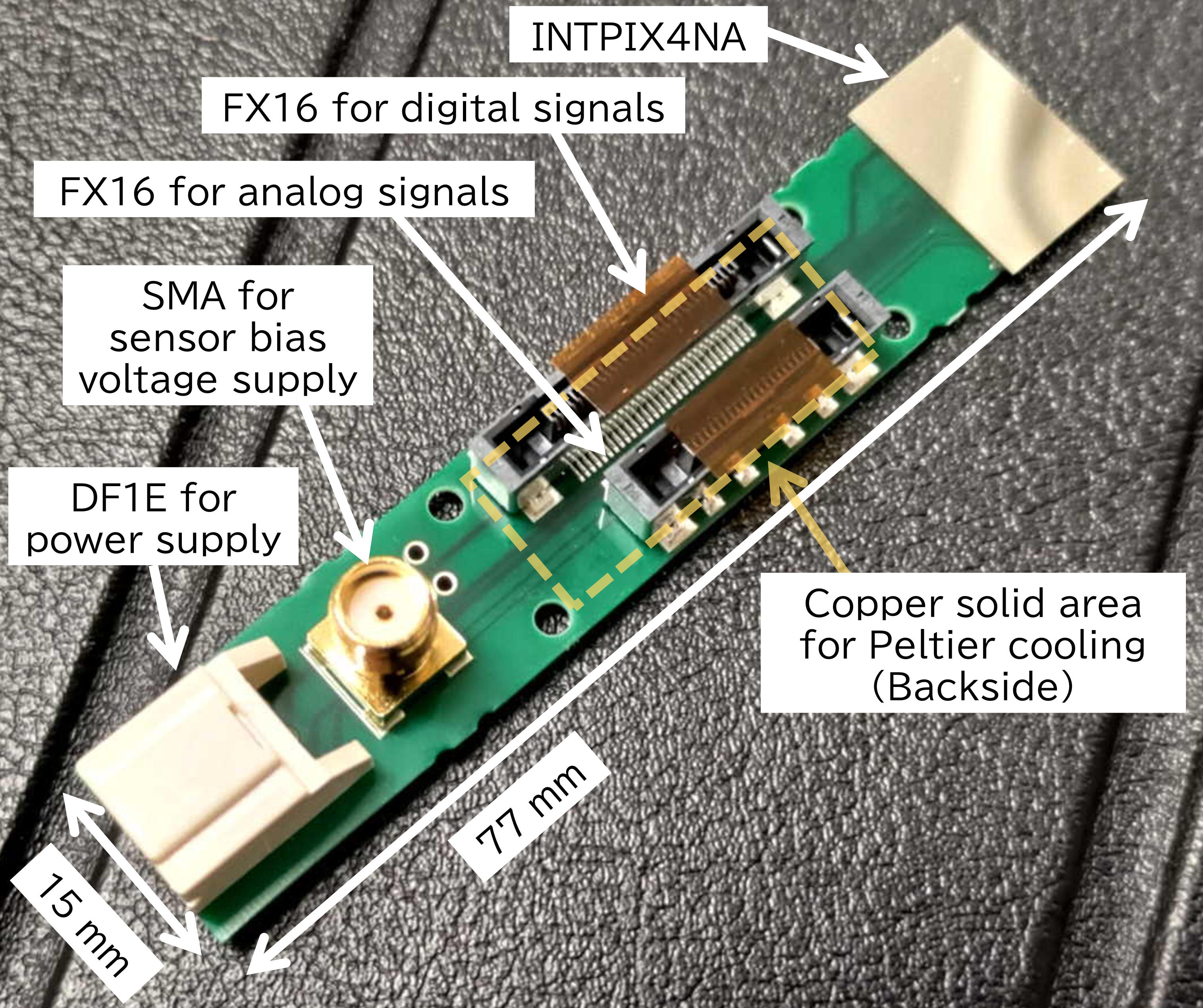}
\caption{Photograph of INTPIX4NA implemented on PCB.}
\label{fig:intpix4na-on-pcb}
\end{figure}

A longer exposure time (\verb|~|0.5 second/frame) is required for X-ray imaging under low X-ray intensity conditions owing to the weak signal. 
However, the INTPIX4NA could not be operated at such long exposure times at room temperature (approximately 25--30 \si{\degreeCelsius}) because of thermal noise. 
Therefore, we developed a new detector board that uses Hirose FX16 connectors for the interface analog/digital signals, Hirose DF1E connector for the power supply, and SMA connector for the sensor bias voltage supply, and can be cooled using a Peltier cooling system (described later). 
The outer dimensions of the board are 15 $\times$ 77 $\mathrm{mm^2}$. 
A 13 $\times$ 27 $\mathrm{mm^2}$ copper solid area is placed on the back side of the board (behind the FX16 connectors) for cooling. 
A photograph of INTPIX4NA implemented on PCB is shown in Fig. \ref{fig:intpix4na-on-pcb}. 

\section{PF-DAQSIX readout system and Peltier detector cooling system}

\subsection{Overview of PF-DAQSIX with Peltier detector cooling}

\begin{figure}[htb]
\centering
\includegraphics[width=\linewidth]{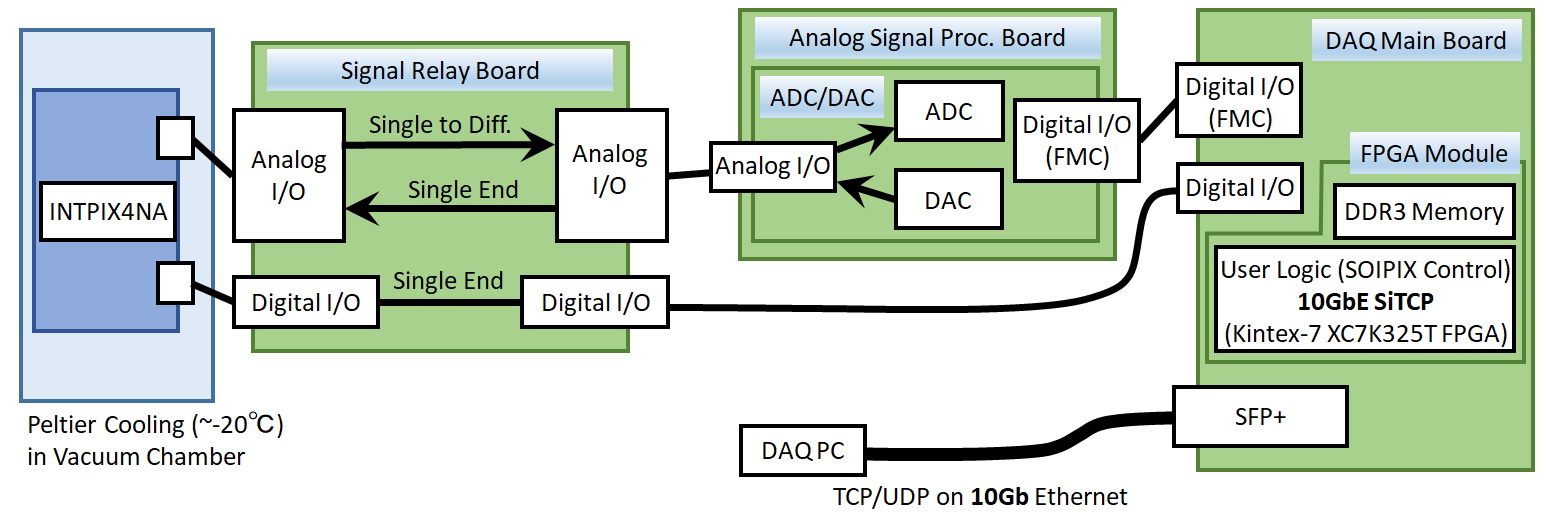}
\caption{Schematic of PF-DAQSIX system with Peltier detector cooling system.}
\label{fig:pf-daqsix-schem}
\end{figure}

The Photon Factory Data Acquisition system for SOIPIX Imaging with XG-Ethernet (PF-DAQSIX) is a readout system developed for the INTPIX4NA SOIPIX detector at KEK-PF. 
This system comprises a KX-Card7, an AMD/Xilinx Kintex-7 FPGA module by Prime Systems, Inc. \cite{psi}, and several interface boards developed by KEK-PF and its collaborators. 
SiTCP-XG \cite{sitcp-xg, sitcp-xg_protodaq} serves as the FPGA-based 10 Gbps Ethernet network controller for high-speed data transfer. 
This readout system was connected to and controlled by a standard PC/AT-based workstation with a 10 Gbps Ethernet interface. 
The INTPIX4NA detector was connected to this readout system via the vacuum feedthrough of the vacuum chamber of the Peltier detector cooling system. 
A simplified schematic is depicted in Fig. \ref{fig:pf-daqsix-schem}. 

\subsection{Main readout boards}

\begin{figure}[htb]
\centering
\includegraphics[width=\linewidth]{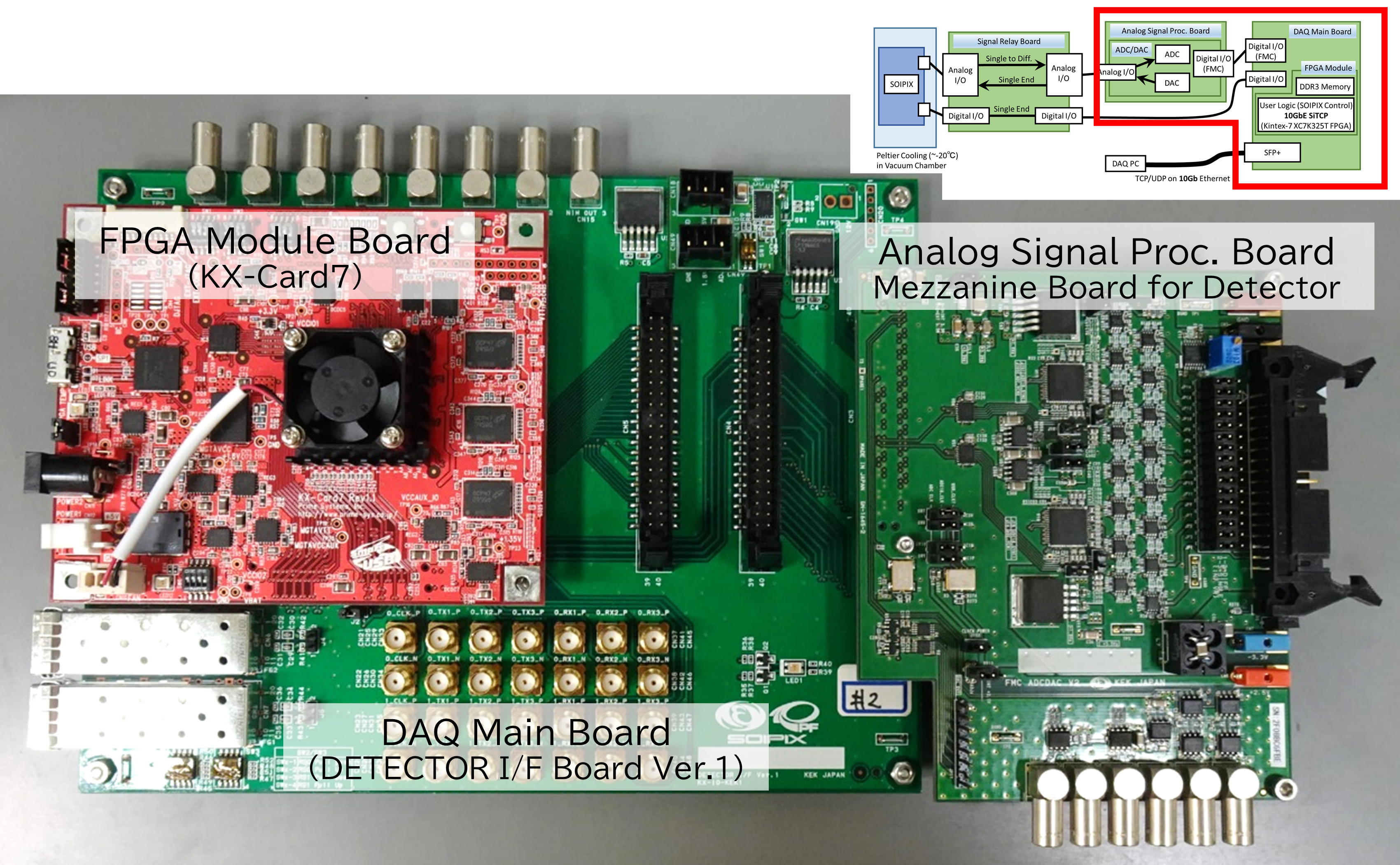}
\caption{Photograph of the main readout boards of PF-DAQSIX system (upper right shows the corresponding part in the schematic).}
\label{fig:pf-daqsix-main}
\end{figure}

The main readout boards comprise an FPGA module board (KX-Card7), a DAQ main board (DETECTOR I/F Board Ver.1 developed by KEK-PF), and an analog signal processing board developed by KEK-PF and collaborators. 
The FPGA module board serves as the core of this readout system. All detector control and data transfer logic, including SiTCP-XG, is implemented in the onboard Kintex-7 FPGA. 
The DAQ main board acts as an interface extension of the FPGA module board and features interfaces for two SFP+, four channels of fast NIM input/output, two channels of MIL-C-83503, and one FMC (ANSI/VITA 57.1) low-pin-count connector. 
The analog signal processing board provides analog-to-digital conversion (ADC) and  digital-to-analog conversion (DAC) functions. 
It is equipped with 16 and 8 channels of 12-bit ADC and 12-bit DAC, respectively. 
A photograph of the main readout board is shown in Fig. \ref{fig:pf-daqsix-main}. 

\subsection{Signal relay board}

\begin{figure}[htb]
\centering
\includegraphics[width=\linewidth]{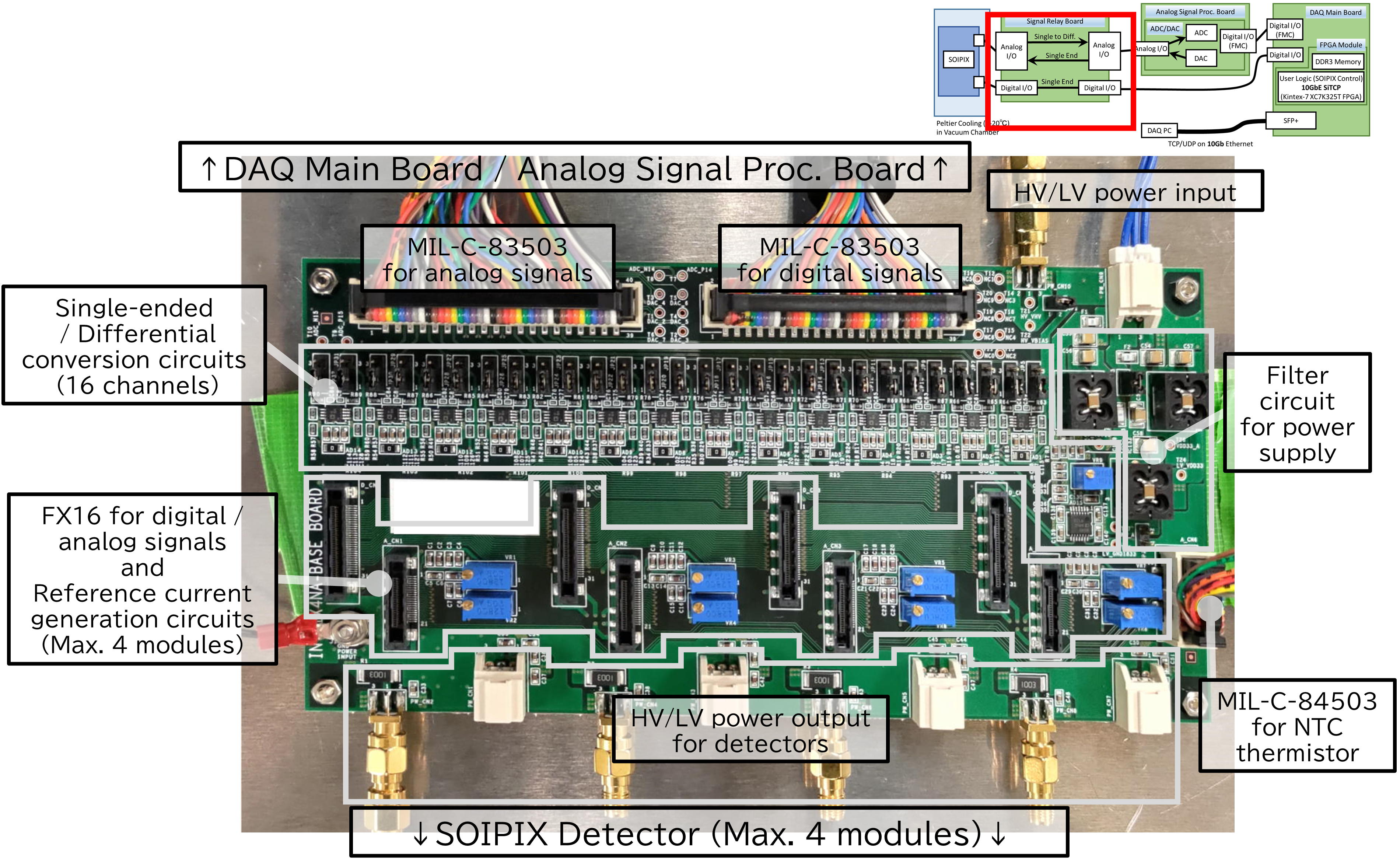}
\caption{Photograph of the signal relay board of PF-DAQSIX system (upper right shows the corresponding part in the schematic).}
\label{fig:pf-daqsix-relay}
\end{figure}

A signal relay board is positioned between the INTPIX4NA detector and the main readout boards. 
The relay board and the detector are linked by 1-2 m ultrafine coaxial cables with a Hirose FX16 connector, a power cable with a Hirose DF1E connector, and an SMA cable for the sensor bias voltage. 
The board and the main readout boards are connected by 1-2 m of flat cables with an MIL-C-83503 connector and a power cable with a Hirose DF1E connector. 
This board relays power, digital signals, and reference voltages. 
Additionally, it generates a reference current, and performs single-ended/differential conversion of analog signals from the detector. 
This board can accommodate connections with up to 4 detectors. 
A photograph of the signal-relay board is shown in Fig. \ref{fig:pf-daqsix-relay}. 

\subsection{Peltier detector cooling system}

\begin{figure}[htb]
\centering
\includegraphics[width=\linewidth]{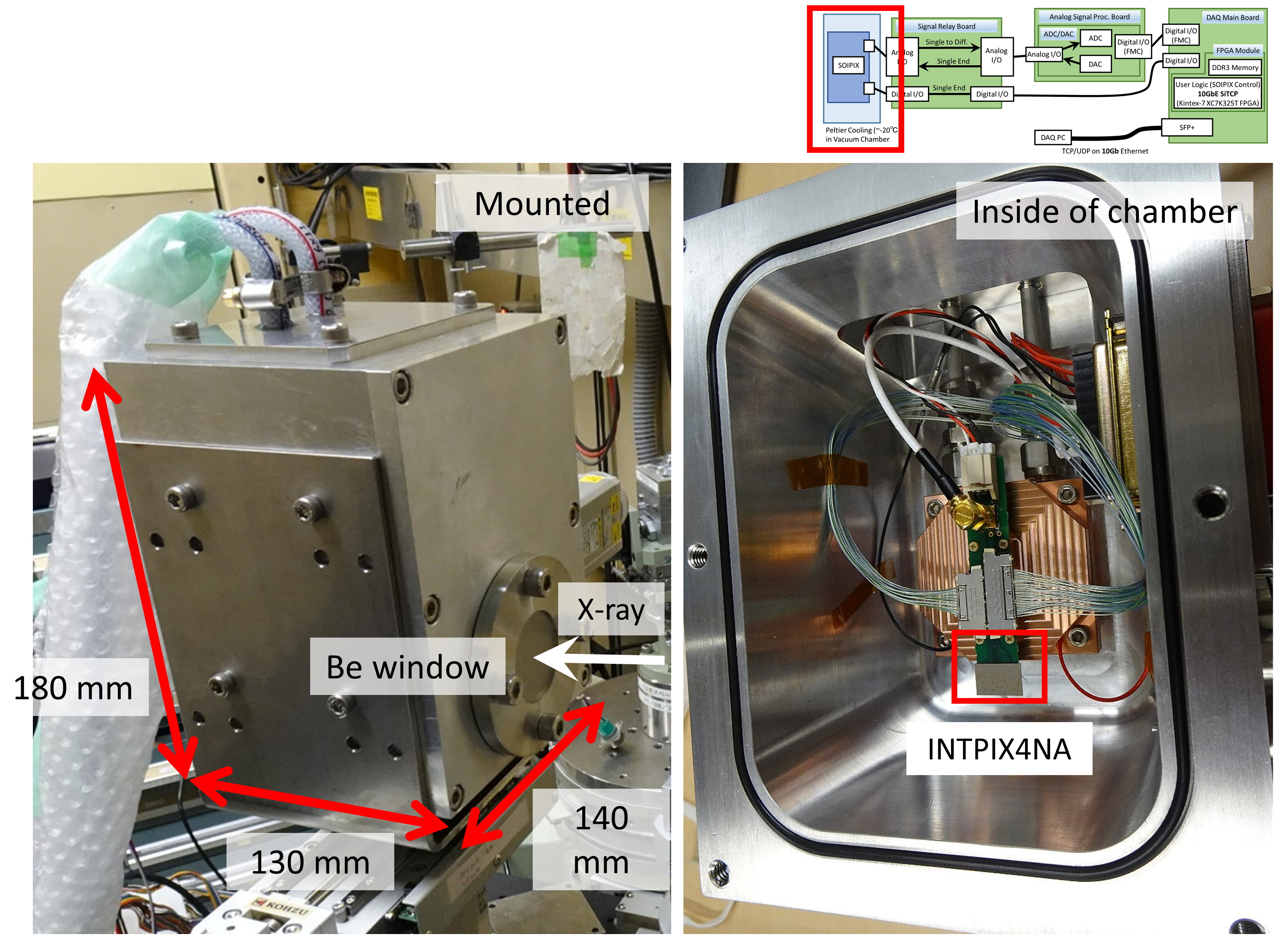}
\caption{Photograph of Peltier detector cooling system (upper left shows corresponding part in schematic).}
\label{fig:peliter-chamber}
\end{figure}

The Peltier detector cooling system comprises an aluminum vacuum chamber (140 mm $\times$ 180 mm $\times$ 130 mm), a Peltier cooling unit, water-cooling pipes for exhaust heat, a beryllium window ($\phi$ 35 mm $\times$ 0.25 mm) for X-ray injection, and a vacuum feedthrough for electrical wiring. 
The Peltier cooling unit is in contact with the water-cooled pipes and the INTPIX4NA detector via copper fixtures. 
The INTPIX4NA is mounted on the copper fixture. 
The signals and power lines of INTPIX4NA are connected to the signal relay board via the vacuum feedthrough. 
During the cooling process, the chamber was first vacuumed for heat insulation and to avoid condensation before cooling the INTPIX4NA to -15 to -20 \si{\degreeCelsius} (typical). 
Detector temperature is monitored using an NTC thermistor (Murata NCP18WB473F10RB \cite{murata-ntc}) implemented on the INTPIX4NA detector PCB board and is feedback controlled. 
Waste heat transported by the Peltier unit from the detector is removed from the chamber using the water-cooling system. 
Therefore, INTPIX4NA can be operated with a longer exposure time (0.5 s/frame). 
A photograph of the cooling system with INTPIX4NA is shown in Fig. \ref{fig:peliter-chamber}. 

\section{Experiment}

\subsection{Energy resolution and equivalent noise charge test}

The energy resolution and equivalent noise charge (ENC) were tested at PF BL-14A. 

\subsubsection{Setup}

\begin{figure}[htb]
\centering
\includegraphics[width=\linewidth]{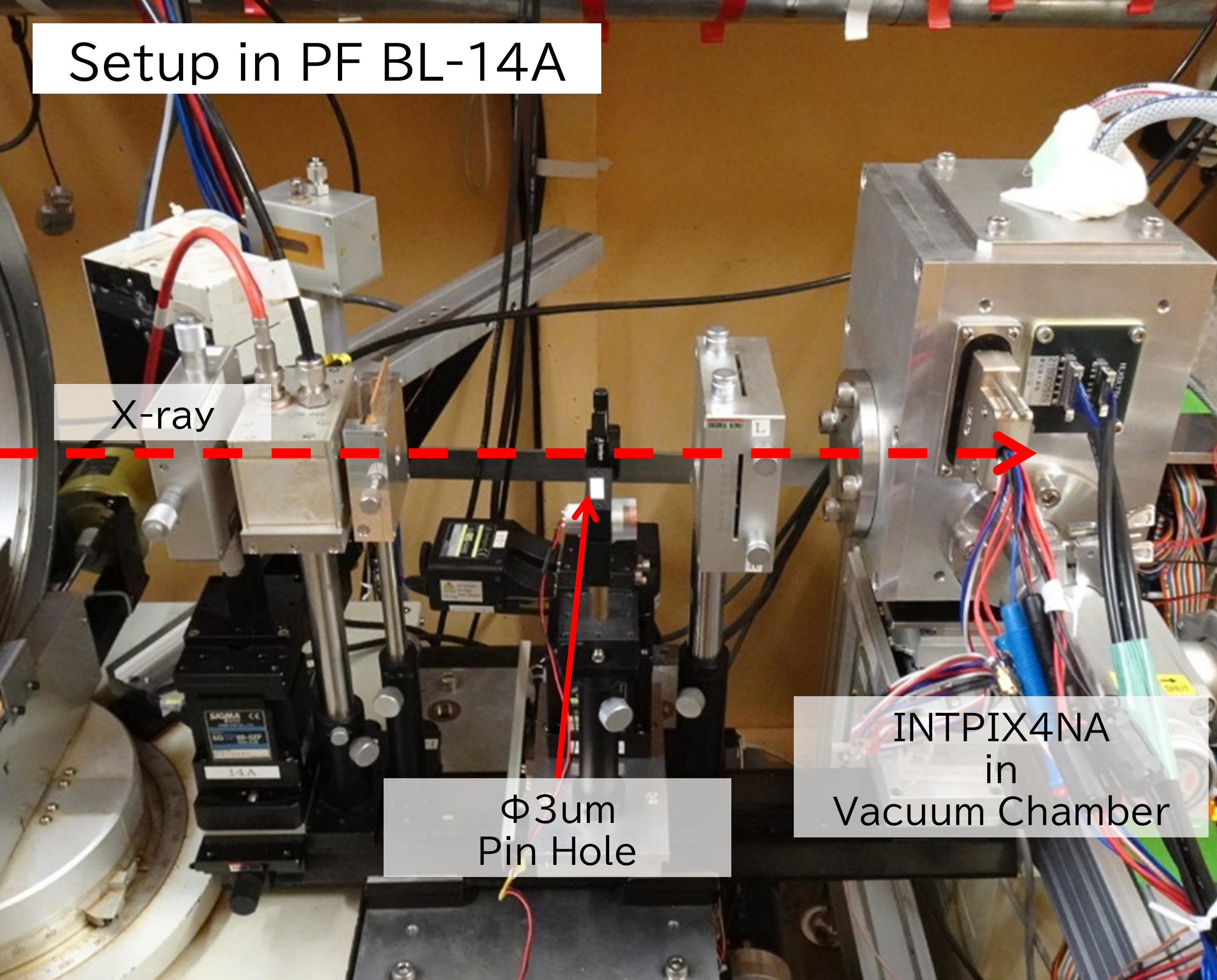}
\caption{Experimental setup of energy resolution and equivalent noise charge (ENC) test at PF BL-14A}
\label{fig:setup-bl-14a}
\end{figure}

An X-ray beam with a monochromatic energy of 12 keV was collimated by a $\phi$ 3 um pinhole and injected into a pixel. 
The detector temperature was -20 \si{\degreeCelsius} (feedback controlled using NTC thermistor) and the exposure time was 1 ms $\times$ 9995 frames. 
The energy resolution ($R(\%) = FWHM_{peak}/Mean_{peak} \times 100$) was derived from the distribution of the total analog signal from a 3 $\times$ 3 matrix, including the pixel of the injected beam and the adjacent 8 pixels. 
The ENC per pixel was calculated from the standard deviation of the background peak of the distribution and the pixel gain (the typical value is 10 \si{\micro V} / \si{\elementarycharge} \cite{i4na}). 
A photograph of the setup is shown in Fig. \ref{fig:setup-bl-14a}. 

\subsubsection{Results and Discussion}

\begin{figure}[htb]
\centering
\includegraphics[width=\linewidth]{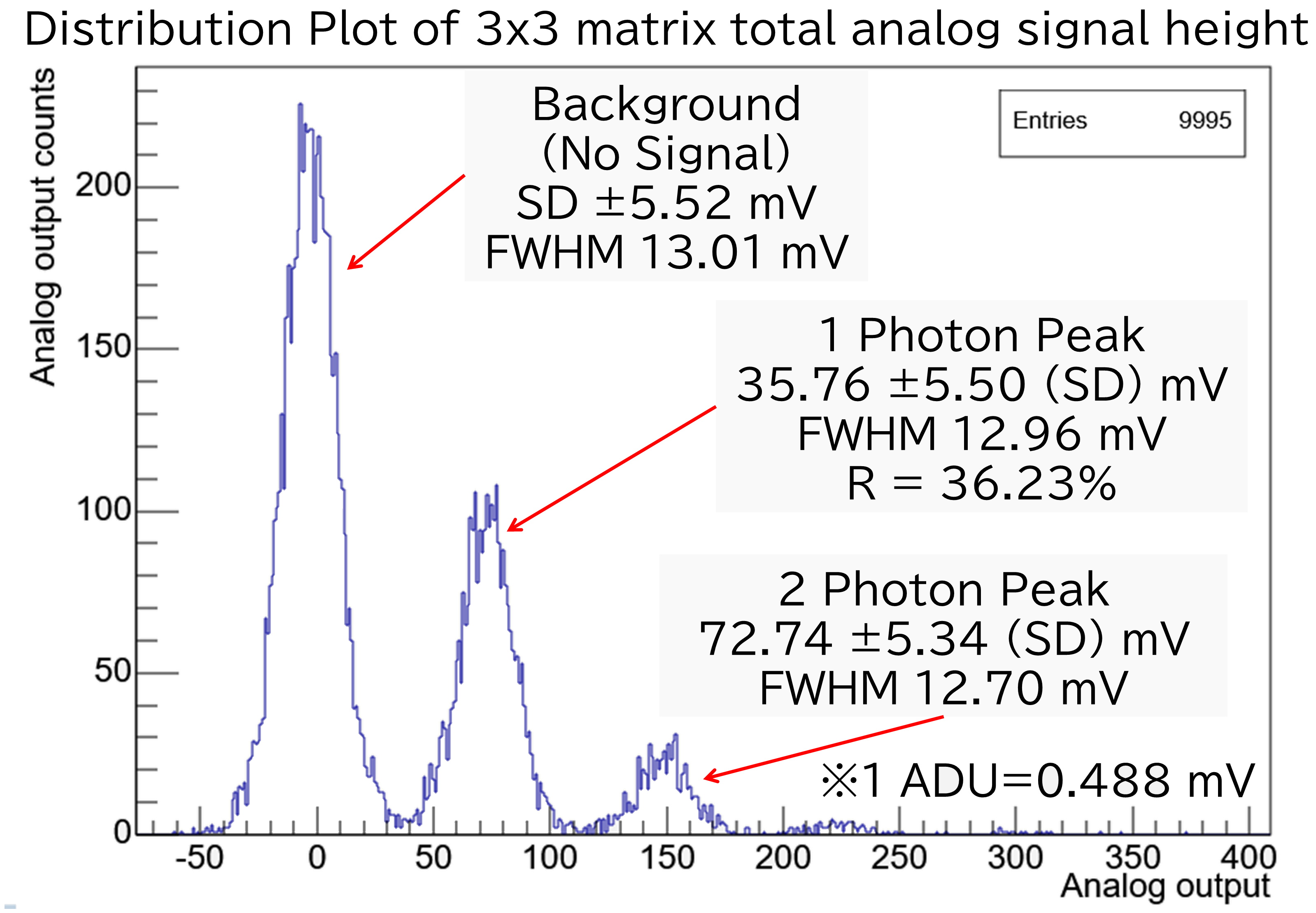}
\caption{Result of energy resolution and ENC test at PF BL-14A}
\label{fig:result-bl-14a}
\end{figure}

Fig. \ref{fig:result-bl-14a} presents the distribution of the total analog signal from a 3 $\times$ 3 matrix, including the pixel (Col 415, Row 279) and the adjacent 8 pixels. 
Based on the measurement results of a previous study \cite{i4na} and these measurements, there is no clear difference in the ENC by the position of the 3 $\times$ 3 matrix on the entire pixel matrix. 
Therefore, we display the 3 $\times$ 3 matrix of the pixel (Col 415, Row 279) and the adjacent 8 pixels as the representative result. 
The energy resolution is 36.23 \%@12 keV, calculated from the peak of one photon in this plot. 
The standard deviations of the background peak are 5.52 and 0.61 mV per pixel. 
Thus, the ENC per pixel is 61.36 \si{\elementarycharge}. 
This is significantly smaller than the charge amount for a 5 keV photon (approximately 1400 \si{\elementarycharge} / photon). 
Therefore, we confirmed that this X-ray camera could separate X-ray signals from noise within the expected range of X-ray energies, that is, 5--20 keV. 

\subsection{Spatial Resolution test}

The spatial resolution was tested at PF BL-14B. 

\subsubsection{Setup}

\begin{figure}[htb]
\centering
\includegraphics[width=\linewidth]{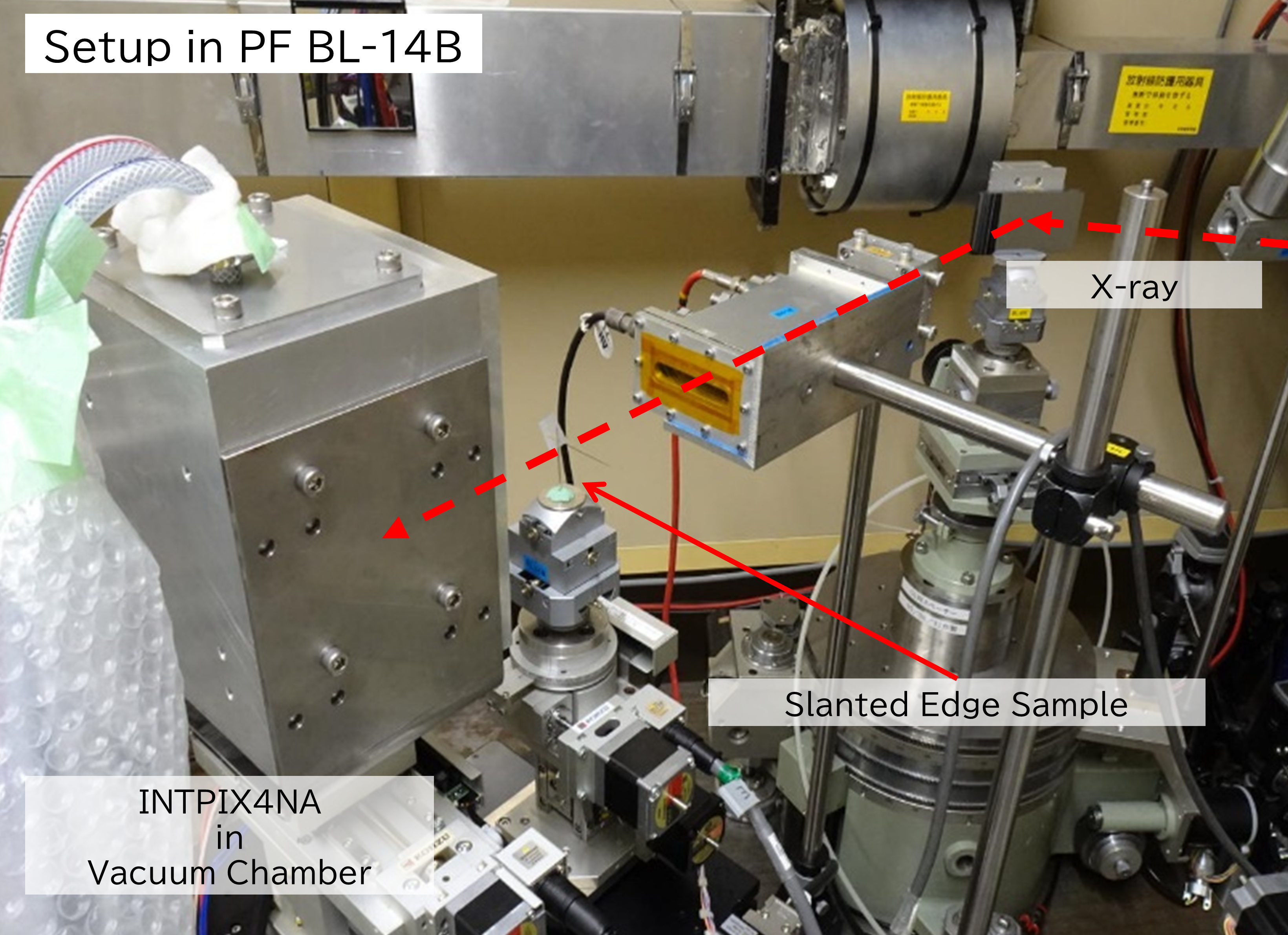}
\caption{Experimental setup of spatial resolution test at PF BL-14B}
\label{fig:setup-bl-14b}
\end{figure}

The monochromatic X-ray energy used was 9.6 keV. The detector temperature was -20 \si{\degreeCelsius} (feedback controlled using NTC thermistor), and the exposure time was 1 s (1 ms $\times$ 1000 frames). 
This test evaluated the spatial resolution as a modulation transfer function (MTF) and was measured using the slanted-edge method \cite{mtf}. 
A photograph of the setup is shown in Fig. \ref{fig:setup-bl-14b}. 

\subsubsection{Results and Discussion}

\begin{figure}[htb]
\centering
\includegraphics[width=0.65\linewidth]{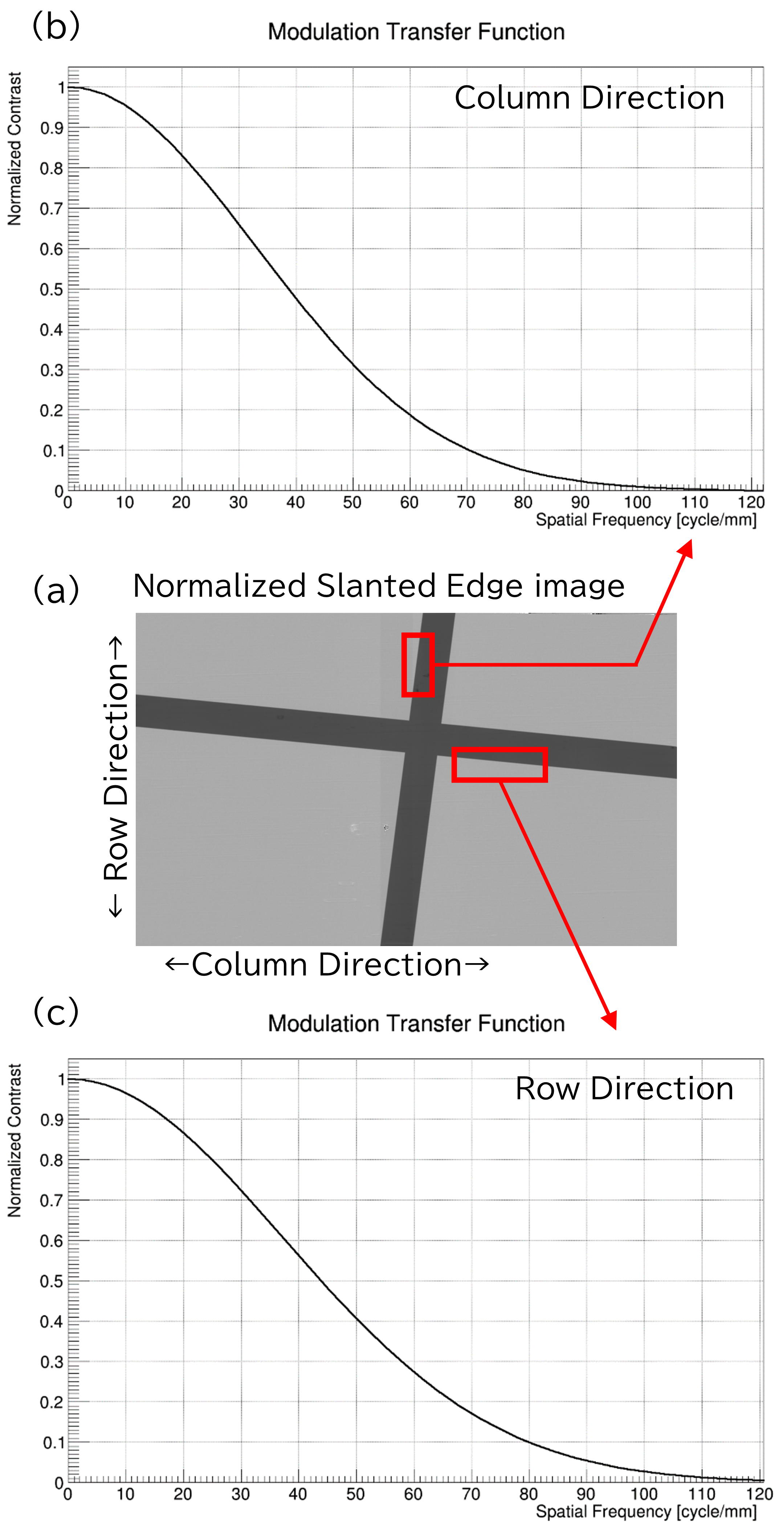}
\caption{Results of the spatial resolution test at PF BL-14B. (a) The normalized slanted edge captured by INTPIX4NA. (b) Modulation transfer function (MTF) in the detector's column direction calculated from the upper area enclosed by the solid line shown in (a). (c) MTF in the detector's row direction calculated from the lower area enclosed by the solid line shown in (a). }
\label{fig:result-bl-14b}
\end{figure}

Fig. \ref{fig:result-bl-14b} shows the results of the spatial resolution test. 
Fig. \ref{fig:result-bl-14b} (a) shows the normalized slanted edge taken by INTPIX4NA. 
Fig. \ref{fig:result-bl-14b} (b) shows modulation transfer function (MTF) in the detector's column direction calculated from the upper area enclosed by solid line shown in (a). 
Fig. \ref{fig:result-bl-14b} (c) shows MTF in the row direction calculated from the lower area enclosed by solid line shown in (a). 
The Nyquist frequency of this detector is 29.4 cycle/mm and the MTF is over 65 \%. 
This demonstrates sufficient performance and is consistent with the prior evaluation results of the INTPIX4NA detector \cite{i4na}. 
Therefore, we confirmed the successful packaging of the INTPIX4NA as an X-ray camera without compromising its spatial resolution performance. 

\subsection{Imaging with low-intensity X-ray}
Low-intensity X-ray imaging was performed at PF AR-NE1A. 

\subsubsection{Setup}

\begin{figure}[htb]
\centering
\includegraphics[width=\linewidth]{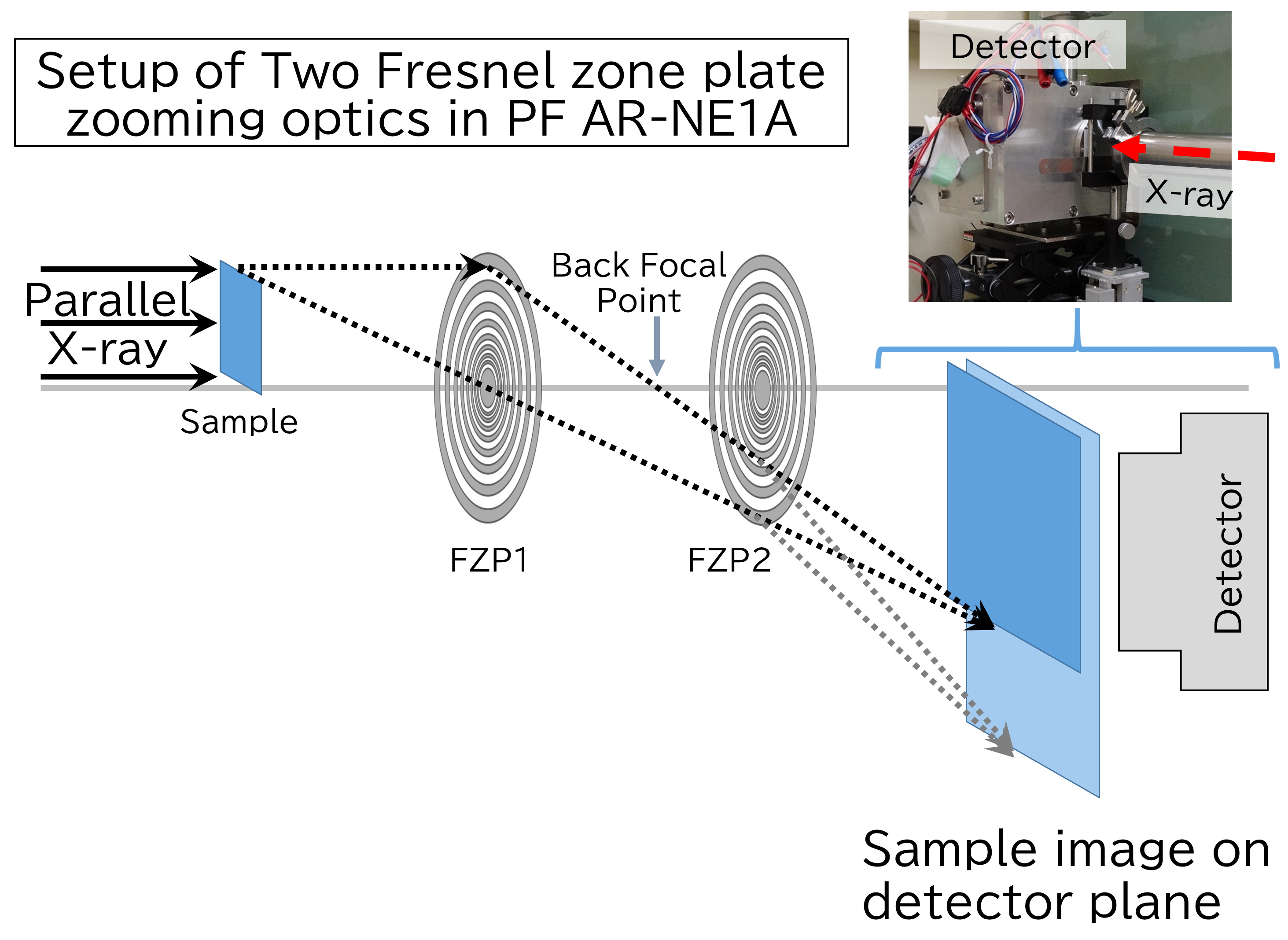}
\caption{Experimental setup of imaging with low-intensity X-ray at PF AR-NE1A}
\label{fig:setup-ar-ne1a}
\end{figure}

The setup schematic is presented in Fig. \ref{fig:setup-ar-ne1a}. 
The X-ray energy is 14.4 keV monochromatic. 
The sample image was magnified using two FZP zooming optics \cite{schmic1} with a magnitude of 115.44. 
A test pattern (developed by Applied Nanotools) was used as the test sample. 
The test pattern image was captured using the INTPIX4NA (at a temperature of -20 \si{\degreeCelsius}, feedback controlled using NTC thermistor) and a scientific CMOS (sCMOS) camera (at room temperature, without any temperature control). 
The sCMOS camera is Hamamatsu Photonics C12849-111U \cite{xscmos1} (2048 \si{\times} 2048 pixel matrix, 6.5 \si{\times} 6.5 \si{\micro\meter^2} pixel, 10 \si{\micro\meter} thickness GOS (P43) scintillator).
In this experiment, the total exposure time is 500 s. 
With INTPIX4NA, this was acquired by integrating 0.5 sec / frame \si{\times} 1000 frames, and with sCMOS, by integrating 5 sec / frame \si{\times} 100 frames. 
The exposure time per frame is set to the longest time within the range where saturation because of thermal noise does not occur in each detector. 
Additionally, the cooling of the sCMOS camera is not as aggressive as that of the INTPIX4NA because this camera package is air-cooled only. 

\subsubsection{Results and Discussion}

\begin{figure}[htb]
\centering
\includegraphics[width=0.7\linewidth]{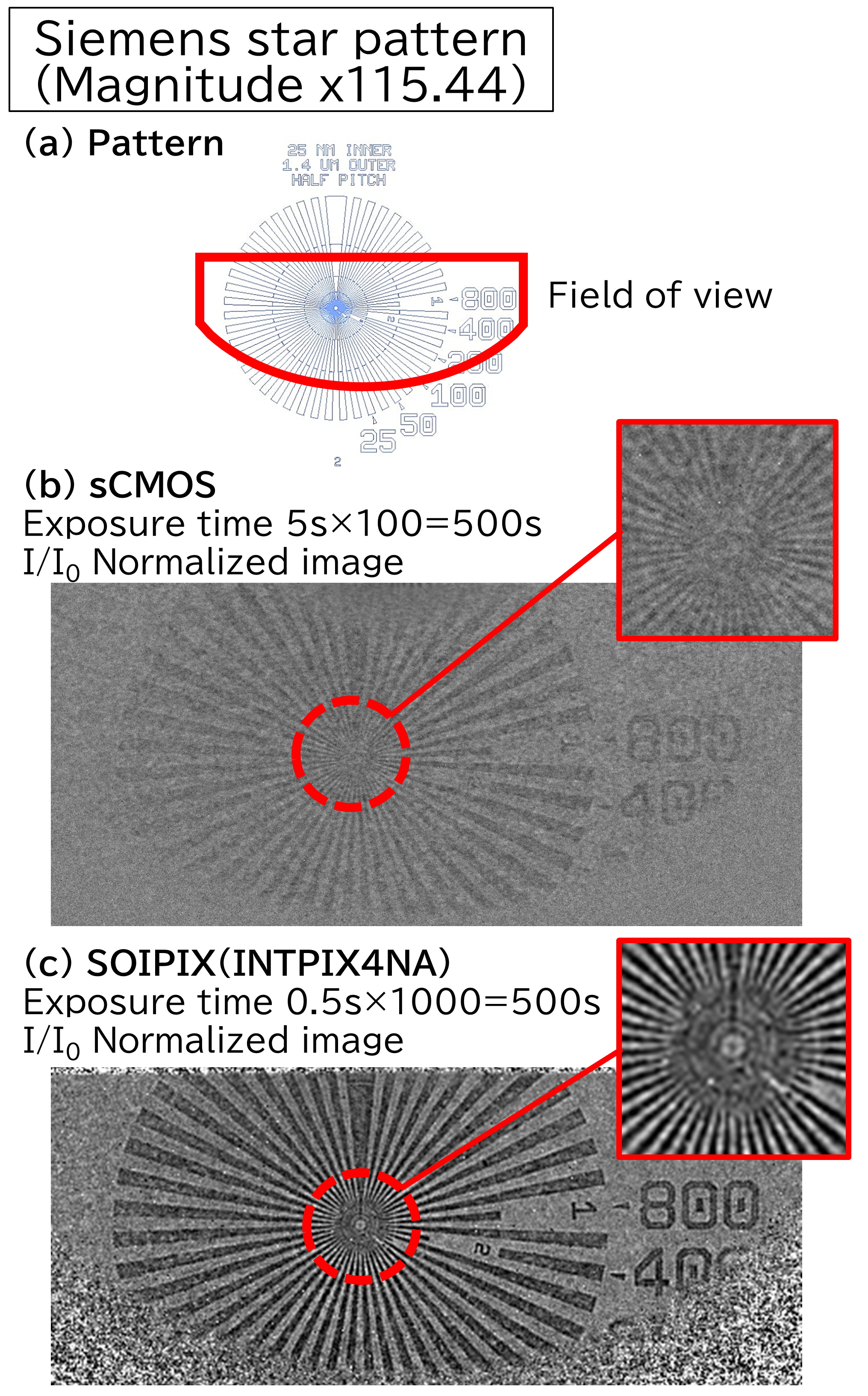}
\caption{Imaging results with low-intensity X-ray at PF AR-NE1A. (a) Pattern of the test chart within the field of view. (b) Normalized image captured using sCMOS. The upper left corner displays an expanded image of the center of the star pattern. (c) Normalized image taken by INTPIX4NA. The upper left corner displays an expanded image of the center of the star pattern.}
\label{fig:result-ar-ne1a}
\end{figure}

Fig. \ref{fig:result-ar-ne1a} shows the test pattern (siemens star) images taken using INTPIX4NA and sCMOS cameras. 
Fig. \ref{fig:result-ar-ne1a} (a) displays the pattern of the test chart with the field of view. 
Fig. \ref{fig:result-ar-ne1a} (b) is the normalized image taken by sCMOS. The upper-left panel shows an expanded image of the center of the star pattern. 
Fig. \ref{fig:result-ar-ne1a} (c) is the normalized image taken by INTPIX4NA. The upper-left panel shows an expanded image of the center of the star pattern. 
The contrast settings in Figs. \ref{fig:result-ar-ne1a} (b) and (c) are the same. 

A comparison of Figs. \ref{fig:result-ar-ne1a} (b) and (c), reveals that the INTPIX4NA image has higher contrast than the sCMOS image. 
In particular, the pattern of the central region, which could not be resolved using sCMOS, is resolved. 

This result indicates that the INTPIX4NA X-ray camera is well-suited for X-ray imaging using 5--20 keV X-rays under low-intensity, low-contrast conditions, such as when capturing soft tissues with poor contrast, objects with fine structures, and specimens vulnerable to radiation damage. 
When imaging specimens under low-intensity, low-contrast conditions using a detector with low resolution characteristics, more X-ray photons are required to compensate for the blur caused by the detector to obtain a sufficiently resolved image. 
Conversely, a detector with high-resolution characteristics can achieve resolution with a minimal number of X-ray photons, enabling imaging with lower intensity X-rays. 
Consequently, it is possible to reduce the X-ray dose needed to obtain the same image quality. 

\section{Conclusions}
This study tested the SOIPIX INTPIX4NA X-ray detector camera developed at KEK-PF. 
The camera was evaluated at KEK-PF experimental stations BL-14A, BL-14B, and AR-NE1A. 
Adequate detection performance was confirmed at PF BL-14A and BL-14B. 
The results demonstrated higher resolution performance than sCMOS under low X-ray intensity conditions using two FZP zooming optics in PF AR-NE1A.
We conclude that this X-ray camera is well-suited for X-ray imaging using 5--20 keV X-rays under low-intensity, low-contrast conditions, such as when capturing soft tissues that do not contrast well, objects with fine structures, and specimens vulnerable to radiation damage. 
In the future, we plan to integrate this X-ray camera into two FZP optics to perform high-resolution X-ray imaging using low-intensity X-rays on practical samples. 
We also plan to apply this camera to X-ray imaging of samples under low-brightness and / or low-contrast conditions, such as biological samples, and to measurement methods for which contrast characteristics are crucial, such as X-ray interferometry. 

\section{Acknowledgements}

This study was approved by the Photon Factory Program Advisory Committee (Proposal Nos. 2021G614 and 2021PF-S001). 
It was supported by JSPS KAKENHI Grant Number 21K14174 and the ISIJ Research Promotion Grant from the Iron and Steel Institute of Japan. 
The study was conducted on behalf of the SOIPIX collaboration \cite{soipix}. 
Certain components of the experimental setup were manufactured by the KEK Mechanical Engineering Center. 
SiTCP-XG was provided by Bee Beans Technologies Co. Ltd. For the latest detailed information on PF-DAQSIX, please refer to the PF-DAQSIX webpages on KEK Wiki \cite{pfdaqsix}. 


\end{document}